# Space weathering of asteroids

D. I. Shestopalov[a], L. F. Golubeva[a], E. A. Cloutis[b]

[a] *Shemakha Astrophysical Observatory, Azerbaijan Academy of Science, Shemakha AZ-324, Azerbaijan;*

[b] *Department of Geography, University of Winnipeg, 515 Portage Avenue, Winnipeg, MB, Canada R3B 2E9*

**Abstract**

Analysis of laboratory experiments simulating space weathering optical effects on atmosphereless planetary bodies reveals that the time needed to alter the spectrum of an ordinary chondrite meteorite to resemble the overall spectral shape and slope of an S-type asteroid is about ~ $10^5$ yr. The time required to reduce the visible albedo of samples to ~ 0.05 is ~ $10^6$ yr. Since both these timescales are much less than the average collisional lifetime of asteroids larger than several kilometers in size, numerous low-albedo asteroids having reddish spectra with subdued absorption bands should be observed instead of an S-type dominated population. It is not the case because asteroid surfaces cannot be considered as undisturbed, unlike laboratory samples. We have estimated the number of collisions occurring in the time of $10^5$ yr between asteroids and projectiles of various sizes and show that impact-activated motions of regolith particles counteract the progress of optical maturation of asteroid surfaces. Continual rejuvenation of asteroid surfaces by impacts does not allow bodies with the ordinary chondrite composition to be masked among S asteroids. Spectroscopic analysis, using relatively invariant spectral parameters, such as band centers and band area ratios, can determine whether the surface of an S asteroid has chondritic composition or not. Differences in the environment of the main asteroid belt versus that at 1 AU, and the physical difference between the Moon and main belt asteroids (i.e., size) can account for the lack of lunar-type weathering on main belt asteroids.

*Keywords*: Asteroids, surface, space weathering, composition; Spectroscopy

# 1. Introduction

Space weathering is a notion defining physical, chemical, and morphological alterations of planetary surface not covered by an atmosphere under the action of exogenous factors such as meteorite bombardment, temperature variations, solar wind particles, and galactic cosmic rays. Solar wind sputtering and micrometeorite impact vaporization generate thin vapor-deposited coatings containing submicroscopic spherules of reduced iron ($Fe^0$) on regolith particles. In spite of the trace concentration of the $Fe^0$ in regolith particles, nanometer-size iron particles lead to optical maturation of the lunar surface, i.e. to a decrease in its albedo and reddening of spectral slope. The latter is a brief summary of more than twenty years of research carried out by B. Hapke and his colleagues to find answers to the questions why the lunar surface material is so dark and red, and why lunar reflectance spectra contain no strong absorption bands (Hapke, 2001 and references therein). Their finding was verified by transmission electron microscope images which provided direct visual proof of the $Fe^0$-rich coatings on lunar regoliths particles (Keller and McKay, 1997; Wentworth et al., 1999; Pieters et al., 2000). The mathematical theories developed by Shkuratov et al., (1999) and Hapke (2001) quantitatively describe the optical effects of the reduced iron particles and have been applied to laboratory experiments simulating space weathering (Hapke, 2001; Shestopalov and Sasaki, 2003; Starukhina and Shkuratov, 2011; Lucey and Riner, 2011) and to the regoliths of atmosphereless cosmic bodies (e.g., Shkuratov et al., 1999; Starukhina and Shkuratov, 2001; Shestopalov and Golubeva, 2004). Recently Noble et al. (2007) founded that the VNIR reflectance spectra of laboratory analogs imitating planetary soils depend significantly not only on concentration of the reduced iron in various samples but also on a size of the iron particles. Larger iron particles (>~ 200 nm) cause an overall decrease of spectral albedo and modest modification of spectral curves while the small iron particles (<~ 40 nm) are responsible for darkening of the surface and its strong spectral reddening. A few years later Starukhina and Shkuratov (2011) and Lucey and Riner (2011) calculated this grain size effect mathematically using Mie theory together with the scattering theories referred to above. Hörz et al. (2005) demonstrated experimentally that impacts at ~ 5 km/s lead to melting of ordinary chondrite regolith and to dissemination of pre-existing metals and sulfides. Such metal droplets (< 100 nm in size), disseminated throughout silicate glass, are supposed to produce space weathering effect on S-type asteroids if they, of course, have chondritic composition. Unfortunately, the absence of spectral measurements of altered chondritic samples in this investigation does not allow us to estimate this optical effect.

There is no doubt that space weathering is an universal process widespread in the Solar System but the intensity of the process and, consequently, its optical effect depend on the heliocentric distance, and the chemical and physical properties of planetary surfaces. The conclusion that asteroid surfaces are optically less mature than lunar soil was made at the end of the last century (Matson et al., 1977; Golubeva et al., 1980) and has been repeatedly corroborated (Pieters et al., 2000; Hapke, 2001; Golubeva and Shestopalov, 2003; Noble et al., 2007). The simplest evidence of this is the fact that weak absorption bands are seen in visible wavelength region of asteroid spectra whereas these bands have not yet been found in lunar spectra. For example, a narrow spin-forbidden band of iron oxide near 505 nm is detected in spectra of dozens of asteroids of various optical types (Cloutis et al., 2011).

In spite of the low degree of weathering in the asteroid belt, this process could "confuse the cards" when interpreting reflectance spectra of asteroids. Space weathering is often considered as the cornerstone required to resolve the so-called "S-type conundrum", which consists of the apparent scarcity of ordinary chondrite parent bodies among S asteroids (Chapman, 1996, 2004; Pieters et al., 2000; Hapke, 2001). Laboratory experiments simulating micrometeorite impact vaporization and solar wind sputtering show that the spectra of ordinary chondrites and olivinic samples do become similar in general to S- and A-asteroid spectra, respectively (Moroz et al., 1996; Yamada et al., 1999; Sasaki et al., 2001;

Strazzulla et al., 2005; Brunetto et al., 2006a, 2006b; Loeffler et al., 2008, 2009). Based on these experimental results, most researchers tend to a conclusion that most S asteroids have chondritic composition, but which have been modified by space weathering. However, Gaffey (2010) reasonably notes that diagnostic mineralogical parameters, such as absorption band centers and band area ratios that allow mineralogical characterizations of asteroid mafic rocks, are not appreciably affected by asteroid-style space weathering. Since the spectral diagnostic parameters – as is evident from these experiments – are essentially unaffected by laser irradiation or ion beam treatment, Gaffey (2010) deduces that the alteration process fails to explain mineralogically important differences in spectra of ordinary chondrites and S asteroids and so their compositions appear to be different from those of ordinary chondrite meteorites. The theory developed by Shkuratov et al. (1999) was used in our work (Shestopalov and Golubeva, 2004) to simulate optical maturation of ordinary chondrite surfaces. We investigated the variations of spectral characteristics of ordinary chondrites depending on size of host material particles and volume concentration of fine-grained $Fe^0$ and found no matches with the same spectral characteristics of S-type asteroids. Thus we inferred that the spectral differences in question are caused not only by a weathering process but also systematic differences in the material compositions of S asteroids and ordinary chondrites.

The efforts of some investigators have been focused on estimation of a space weathering rate in the asteroid belt (Lazzarin et al,. 2006; Marchi et al., 2006a, 2006b; Paolicchi et al., 2009, 2007; Nesvorný et al., 2005; Willman et al., 2008; Willman and Jedicke, 2011). A principal motivating force of these studies was to find a rational explanation of the paucity of spectroscopic analogues of ordinary chondrite meteorites among silicate-rich S-type asteroids. If space weathering does take place in the main belt, then spectra of relatively young S asteroids (small near-Earth objects or the members of dynamically young families) should differ from those of relatively old S asteroids (large objects of the main belt or the members of dynamically old families). A spectral slope in the vis-NIR range was chosen as a space weathering index, which was supposed to be directly correlated with the different relative age of asteroids. Despite the fact that this choice cannot be recognized as most suitable for surface maturity characterization (Gaffey, 2010), the final result of the investigations was rather contrary to expectations (Paolicchi et al., 2009; Willman and Jedicke, 2011). It turned out that an astrophysical time scale (i.e. the time necessary to change spectra of, say, small Q-type asteroids to typical spectra of large S-type asteroids) is greater by some orders of magnitude than space weathering timescales estimated from laboratory experiments (e.g., Hapke, 2001; Loeffler et al., 2009; Brunetto and Strazzlla, 2005). As some investigators improved their age-color-size theory for asteroids, the value of asteroid weathering timescale varies directly with the date of publication: $570 \pm 220$ Myr (Willman et al., 2008), $960 \pm 160$ Myr (Willman et al., 2010), $2050 \pm 80$ Myr (Willman and Jedicke, 2011). This implies that the space weathering rate in the asteroid belt is very low or, equivalently, that the degree of weathering of asteroid surfaces is low (see two previous paragraphs). Moreover, Vernazza et al. (2009) deduce: *"The rapid color change that we find implies that color trends seen among asteroids are most probably due to compositional or surface-particle-size properties, rather than to different relative ages."* This conclusion confirms our result concerning the S-asteroid compositional trend along heliocentric and perihelion distances (Golubeva and Shestopalov (2011) and references therein).

Recently, a polarimetric survey of 22 small main-belt asteroids belonging to the Koronis ($\sim 10^9$ yrs old) and Karin ($\sim 6 \times 10^6$ yrs old) dynamical families was realized by Cellino et al. (2010). The chief aim was to compare the optical properties of asteroid surfaces of nearly identical composition but certainly modified by space weathering mechanism due to essentially different surface exposure times. However no differences were found in the polarimetric properties of the asteroids studied. Particularly, geometric albedo derived from the polarimetric data for Koronis and Karin members turned out to be the same in the range of observational errors and closely approximated to the average albedo of S-type asteroids.

We can ask the question: What are the processes that "blur over" the optical manifestation of space weathering on asteroids? The most popular current ideas are: (i) exposure of fresh subsurface material by the effects of tidal forces during close encounters of asteroids with terrestrial planets (Marchi et al., 2006a; Binzel et al., 2010a, 2010b); and (ii) collisional evolution of asteroids, which steadily refreshes their surface (Richardson et al., 2005; Shestopalov and Golubeva, 2008; Paolicchi et al., 2009; Golubeva and Shestopalov, 2011). The first hypothesis is still being debated in literature and refined (Nesvorný et al., 2010); the latter, we believe, is quite reasonable.

The models of asteroid regoliths developed in the 1970s differ in details, but all predict high mobility of regolith layers on small bodies (e.g. Hausen and Wilkening, 1982 and references therein). This occurs for two reasons: the low gravity of asteroids and the high frequency of inter-asteroid collisions that do not lead to asteroid disruptions but effectively garden their surfaces at the regional and body scales. Below we consider the processes which hamper the progress of optical maturation of asteroid-sized bodies. In the first instance, we need to specify an exposure time during which the surface changes its optical properties under space weathering.

## 2. Timescales of optical maturation of undisturbed surface: data of laboratory experiments.

The exposure time necessary to change the optical properties of lunar or asteroidal surfaces by space weathering have repeatedly been derived from laboratory experiments. However these estimations are very variable. For example, Hapke (2001) supposes the time required to convert initially fresh chondritic surface into S-asteroid regolith under the action of solar wind sputtering is not less than 50000 years. Loeffler et al. (2009) infer the weathering time of about 5000 years at 1 AU from He ion irradiation of olivine powder sample. Sasaki et al. (2001), using nano-pulse laser shots at olivine pellet samples, estimated a weathering time of ~ $10^8$ years at 1 AU and ~ $6\times10^8$ years in the asteroid belt; this result was corroborated by Brunetto et al. (2006b). The weathering timescale of asteroid surfaces is supposed to be of the order of, or less than, $10^6$ years (Brunetto et al., 2006a; Vernazza et al., 2009).

To determine the space weathering timescale as accurately as possible, we use the geometrical-optics model developed by Shkuratov et al. (1999). In this model, the albedo of a particulate surface, $A(\lambda)$, is algebraic function of optical density, $\tau(\lambda)$, of some "average" particle of light-scattering material, $A(\lambda)=F_1(e^{-\tau(\lambda)})$, notably that $\tau(\lambda)=F_2[A(\lambda)]$, i.e., the analytical reciprocity of the model permits estimating the wavelength behavior of optical density from spectral albedo. The functions $F_1(\cdot)$ and $F_2(\cdot)$ can be taken from the cited references. If the grains of reduced iron ($Fe^0$) are embedded into the particle rims and the iron grain size is much less than wavelength of incident light, then (Shkuratov et al., 1999; Hapke 2001):

$$\tau(\lambda) = \tau_h(\lambda) + \tau_{Fe}(\lambda) = \alpha_h(\lambda)l + \beta_{Fe}(\lambda)b,$$

where $\tau_h(\lambda)$ and $\tau_{Fe}(\lambda)$ are the optical densities of a host material and reduced iron, $\alpha_h(\lambda)$ and $\beta_{Fe}(\lambda)$ being specific absorption coefficients of the host material and $Fe^0$, respectively; and $l$ is the mean photon path length through the particles. In turn, $b = 2\delta c_1$, where $\delta$ is the thickness of the particle rim and $c_1$ is the volume concentration of $Fe^0$ in this layer. We used Eqs. (11, 12) from Hapke (2001) and optical constants of iron by Johnson and Cristy (1974) to calculate $\beta_{Fe}(\lambda)$; the dimension of the quantity is $\mu m^{-1}$. This absorption coefficient rises steeply towards short wavelengths, and therefore all attributes of the surface optical maturation process become apparent first in the visible range of spectrum.

Based on the works of Morris (1977; 1980) we can suppose that the volume fraction of the fine-grained metallic iron is a linear function of time, i.e., $c_1(t) = c_{1\,max}(t/T)$, where $c_{1\,max}$ is maximal concentration of $Fe^0$ accumulated for the time $T$, when visual albedo of the surface saturates. Because $c_1$, $\delta$, and $l$ parameters are often unknown for laboratory samples

(and certainly for the particles of asteroid surfaces) it makes sense to work with the *b* parameter:

$$b(t) = 2\delta c_{1\max}\left(\frac{t}{T}\right) = b_{\max}\left(\frac{t}{T}\right). \tag{1}$$

In his experiment, Hapke (1973) exposed silicate rock powders to proton irradiation in a high vacuum environment to simulate darkening of lunar surface materials by solar wind. The total irradiation of the samples was measured as the electric charge density (Coulomb $cm^{-2}$). Since the average solar flux is about $3\times10^8$ protons $cm^{-2}$ $s^{-1}$ at 1 AU, Hapke (1973) derived a simple ratio between the charge density and the exposure time by solar proton irradiation: ~5 C $cm^{-2}$ or $10^4$ yr, ~30 C $cm^{-2}$ or $6\times10^4$ yr, and so on. We used this ratio to modify Figure 3 from the work by Hapke (1973) and plotted a graph of visual albedo of the samples versus the exposure time *t*. Then we calculated theoretical relations between albedo and *t* using Eq. (1) and Shkuratov's model.

As is seen from Figure 1a, albedo of the fine-grained samples varies faster with exposure time than that of the coarse-grained ones. Apparently, this occurs due to a more porous surface of the fine-grained samples, so that vapor-deposited coatings on the particle surfaces form faster. After time $T \sim 2\times10^5$ yr, visual albedo of the olivine basalt powders subjected to proton irradiation changes very slowly. Figure 1b illustrates the fact that, all other things being equal, regardless of the initial albedo of unaltered samples, this value approaches ~0.05 for about the same time interval *T*, which can be called saturation time of solar wind darkening at 1 AU. We corroborate the finding of Hapke (1973) that the undisturbed lunar regolith would darken in $\sim10^5$ yr. This result is also supported by numerical simulation of the sputtering of the lunar soil by solar wind protons (Starukhina, 2003) and Monte Carlo computation of complex dynamical evolution of an uppermost layer of the lunar regolith, the particles of which should show the solar wind saturation effect (Borg et al., 1976).

Figure 2 shows alteration of an ordinary chondrite spectrum with increasing volume fraction of fine-grained $Fe^0$ in the uppermost surface of particles of chondritic material. The originally "chondritic" spectrum became similar in overall shape and slope to that of S-type asteroids at $b \approx 0.002 - 0.006$ μm or, equivalently, between $1 \times 10^4$ and $3 \times 10^4$ yr on the above timescale. For the asteroid belt, we can recalculate this spatter – darkening time using the relation derived by Hapke (2001):

$$t_{MB} = t_{1AU} a^2 \sqrt{1-e^2} ,$$

where $a$ and $e$ are the semimajor axis in AU and the eccentricity of an asteroid orbit. For an asteroid in the middle of the main belt with $e = 0.14$, $t_{MB}$ ranges from $7 \times 10^4$ to $2.2 \times 10^5$ yr or $\sim 1.5 \times 10^5$ yr on average.

Laser irradiation experiments that simulate micrometeorite impact vaporization of regolith particles show that the timescale of optical maturation by this process seems to be longer than the time of solar wind darkening (Sasaki et al., 2001; Brunetto et al., 2006). To convert the spectrum of an originally fresh olivine target bombarded by micrometeorites with a diameter of 1 μm and an impact velocity of 20 km/s into an A-type asteroid spectrum requires $\sim 10^8$ yr in space at 1 AU distance (Sasaki et al., 2001). We recalculated this time interval taking into account the cumulative flux of dust particles in the range of masses from $2 \times 10^{-15}$ to $10^{-7}$ g (approximately 0.12 – 46-μm size particles) at 1 AU taken from McBride and Hamilton (2000). The same spectral alteration (i.e. the spectrum of olivine → roughly A-type spectrum) takes $\sim 1.3 \cdot 10^6$ yr assuming the impact velocity to be 20 km s$^{-1}$. The reduction of the time interval occurs mostly due to rarer but heavier dust particles of $\sim 45$ μm in size.

Figure 3 shows our interpretation of the laser irradiation experiment of Sasaki et al., (2001). The 30-mJ irradiation of the olivine pellet at $b \approx 0.002$ μm corresponds to $1.3 \times 10^6$ yr in space at 1 AU; the saturation time for this process is $\sim 2 \times 10^7$ yr, which is greater by two orders of magnitude than that for solar wind darkening at the same distance from the Sun. Visual albedo of the altered olivine samples at $1.3 \times 10^6$ yr is still far from the saturation level of ~ 0.05. The survey of asteroid reflectance spectra included in the studies of Strazzulla et al. (2005), Brunetto et al. (2005, 2006a, 2006b), and Loeffler et al. (2008, 2009) leads to the conclusion that a saturated level of spectral albedo is not reached in their investigations as well. So, the experiment of Hapke (1973) suggesting short times scales for darkening remains appears to be valid.

It is difficult to recalculate correctly the timescale for microparticle impact darkening in the case of the main belt asteroids since neither a size frequency distribution for microparticles nor their impact velocity are well known there. For example, to evaporate a unit mass of forsterite, an impact velocity of 10 km/s or higher is required (Hapke, 2001 and reference therein). The average relative velocity of asteroids is only ~ 5 km/s and only about 3% of all impacts in the main belt occur at velocities over 10 km/s (O'Brien et al., 2011). The sources of the high-velocity dust particles in the main belt are also unknown today. It appears that the role of microparticle impacts into main belt asteroids is reduced to the crushing of small regolith particles and, possibly, to their partial vitrification.

Thus we conclude that solar wind is apparently the sole and constantly operative process which could lead to optical maturation of asteroid surfaces. We determine that the undisturbed surface of a body having ordinary chondrite composition will acquire the spectral features of an S-type asteroid in $\sim 1.5 \times 10^5$ yr. Furthermore, since the surface maturation of such an asteroid will extend beyond this time, its visual albedo will be reduced to ~ 0.05 by the saturation time of ~ $1.4 \times 10^6$ yr and its reflectance spectrum will become reddish, akin to curve (6) in Figure 2. Both these time intervals are much shorter than a collisional lifetime of asteroids larger than 8 km in diameter (~ $5 \times 10^9$ yr, O'Brien et al., 2005). Consequently, instead of the S-type population we should observe numerous low-albedo asteroids having reddish spectra with subdued absorption bands. This does not happen, of course. The key words for solving this logical contradiction are "undisturbed surface". Such an approach is reasonably sufficient for the Moon but not for asteroids.

As is evident from simulation of the exposure history of lunar regolith particles (Borg et al., 1976 and references therein), the lifetime of a freshly deposited layer before burial by another layer is ~$2.5\times10^7$ yr. As a result of random walks under meteorite impacts, some particles sometimes reach the uppermost layer of the surface, while others descend. The average exposure time of any surface particle to the solar wind is ~ 5000 yr; and the number of exposures of a particle to the solar wind is ~ 20 – 30, so the total exposure time of mature particles is ~ $10^5 - 1.5\times10^5$ yr. The thickness of the uppermost layer of the lunar regolith, consisting of the optically matured particles, is ~ 5 mm. This "skin" (the term by Borg et al., 1976) almost everywhere covers the lunar surface. Largely due to this circumstance, the Moon has low albedo and reddish spectrum. Thus the concept of "undisturbed surface" is quite applicable to the Moon because of the low frequency of meteorite impacts and the short timescale of solar wind darkening. A different situation occurs in the case of asteroids. The timescale necessary to convert a "chondritic" spectrum into S-type is approximately the same as for lunar mature regolith (i.e., ~$10^5$ yr). However, the low gravity of asteroids and the high frequency of non-catastrophic collisional events result in an essential difference in intrinsic mobility of asteroid and lunar regoliths.

Below we examine the main traits of the evolution of asteroid regoliths within the framework of the problem. It should be stressed that our intent here is not a detailed simulation of the asteroid regolith formation but to illustrate the fact that the continuous collisional resurfacing of asteroids hampers strong optical alteration of their surfaces by solar wind.

## 3. Collisions between asteroids

The spectrum of an initially fresh surface exposed to solar wind and having ordinary chondrite composition will exhibit some resemblance to the spectra of S-type asteroids in the time of $T_{OC\to S} \approx 1.5\times10^5$ yr at a distance of about 2.6 AU. At the same time, the surface of the asteroid undergoes numerous collisions with bodies of different masses and sizes. What happens to asteroid surfaces during this relatively short time span?

*3.1. The number and frequency of collisions*

The number of collisions occurring between a target asteroid with diameter $D_a$ and projectiles with diameters equaled to or larger than $D_p$ during a time interval $\tau$ can be estimated by the following equation (e.g., Wetherill, 1967; Farinella et al., 1982):

$$N_p = 0.25 P_i D_a^2 N(>D_p)\tau, \qquad (2)$$

where $P_i$ is the intrinsic collision probability, $N(>D_p)$ is the cumulative number of bodies larger than $D_p$ in the projectile population. All bodies are assumed to be spherical and their diameters are measured in km. If $N_p$ equals 1 in Eq. (2) then $\tau$ is the mean time between impacts; the mean frequency of the impacts is simply expressed in terms of $T_{OC \to S}$, the time span of interest:

$$\frac{1}{\tau_p} = \frac{N_p}{T_{OC \to S}}. \qquad (3)$$

$P_i$ may be interpreted as the collisional rate between two asteroids for which the sum of their radii is 1 km. We are interested in non catastrophic collisions of asteroids, so the diameter of target asteroids is much larger than that of projectiles and a component including $D_p^2$ is omitted in Eq. (2). Calculation of $P_i$ is a non-trivial task for the elliptic and noncoplanar orbits of asteroids. Beginning with the pioneering work of Öpik (1951) and Wetherill, (1967) the intrinsic collision probability for various populations of asteroids both in the main belt and beyond its bounds has been estimated in many studies (Davis et al., 2003). In our work the value for the main belt asteroids was taken from Bottke and Greenberg (1993), $P_i = 2.86 \times 10^{-18}$ km$^{-2}$ yr$^{-1}$.

To calculate $N(>D_p)$ we used an incremental size distribution for projectile population (O'Brien and Greenberg, 2005) derived from modeling collisional and dynamic evolution of the main-belt and near-Earth asteroids. This model size distribution includes asteroids up to 0.001 km in size and was tested against a number of observational constraints such as the observed population of the main belt asteroids and NEAs from various surveys, the cosmic ray exposure ages of meteorites, the cratering records on asteroids.

Figure 4a illustrates the number of asteroid collisions in the time $T_{OC \to S}$ depending on the diameter of projectiles. This is, in fact, the asteroid size frequency distribution (O'Brien and Greenberg, 2005) inverted to $N_p$ using Eq. (2). The model does not indicate a deficit of small asteroids in the main belt therefore we expanded our calculation up to impactor diameters of 0.0001 km. The diameter of the largest non-catastrophic impactor is defined by both the time period (i.e., $T_{OC \to S}$ in the given case) and the asteroid cross section, and varies from 0.0045 to 0.045 km for target asteroids from 20 to 500 km in size. As is seen from Figure 4a, even in this relatively short time span, asteroids undergo a great variety of impacts, which, of course, do not lead to total disruption of asteroid bodies, but effectively garden their surfaces.

Figure 4b demonstrates an essential difference in a "bombardment hazard" in the asteroid belt and in the lunar orbit. For that we used an average lunar impact probability of $1.86 \times 10^{-10}$ yr$^{-1}$ per asteroid as derived by Werner et al. (2002) for all known near-Earth asteroids with absolute magnitude $H \leq 18^m$. To calculate the number of collisions between a 500-km body and projectiles in the lunar orbit we must take into account, firstly, the difference in the lunar and body cross section (i.e., factor of $D_B^2 / D_{MOON}^2$) and secondly, the fact that the cumulative size-frequency distribution of projectile population near 1 AU (Werner et al., 2002) was scaled to $N(>D_p) = 1$ at $D_p = 1$ km. Fortunately, since $N(> 1\ km)$ is known and equals $\sim 700$, we have finally:

$$N_p\big|_{at1AU} = 1.86 \times 10^{-10} \times 700 \times \frac{D_B^2}{D_{MOON}^2} \times N(>D_p) \times T_{OC \to S},$$

where $N(>D_p)$ is the cumulative number of projectiles near Earth. In the model used, the smallest projectile diameter of the near-Earth population is only $10^{-2}$ km thus we were forced to extrapolate our calculations in the range of projectile diameters typical of the asteroid belt. The dotted line in Figure 4b is the result of such an extrapolation. We can not state that the forecast outcome for a Vesta-size body in a lunar orbit is sufficiently precise. However it is clear that the bombardment hazard is much less in the lunar orbit than that in the asteroid belt. This is the most essential condition for the retention of a thin weathered layer on the lunar surface.

The time interval $\tau_p$ between impacts as a function of the projectile diameter for the asteroids of various sizes is plotted in Figure 4c. As is seen from this Figure, the larger the target asteroid the shorter $\tau_p$; the smaller the size of the impactors the higher the frequency of collisions. We can estimate an expected time of collision, $\tau_e$, of an asteroid with any body of the target population. For the given target asteroid, it is simply $T_{OC \to S}$ divided by the total number of impacts, $N_{coll} = \sum N_p$, in this time period. In accordance with Eq. (3)

$$\tau_e = \frac{1}{\sum \left(\frac{1}{\tau_p}\right)}, \qquad (4)$$

where the index $p$ denotes the projectile population with diameter of $D_p$. The $\tau_e$ dependence on asteroid diameter is shown in Figure 4d. A body of 20 km in size may undergo one impact in the expected time of about 210 days; the 50-, 100-, 300-, and 500-km asteroids do in ~ 33 days, 8 days, 22, and 8 hours, respectively. Therefore, asteroid surfaces cannot be considered as undisturbed, they rather are "shock-activated". A heavy shower of impactors generates craters of various sizes on asteroids and triggers impact-induced seismic processes.

*3.2. Craters and crater ejecta*

We will assume for the moment that all impacts into asteroids produce craters. In reality this may not be the case, because some projectiles can move along trajectories nearly tangential to the impact site and perhaps create grooves, but not craters in the usual sense of the term. However, even if the number of such very oblique impacts will be one-half of the total impacts, it has little effect on the order of magnitudes which we utilize in this work.

One of the approaches to solving the tasks of the impact dynamics is scaling laws. This phenomenological approach requires some explanation (Housen et al., 1983; Holsapple, 1993 and reference therein). The scaling laws determine the functional relations between the outcomes of hypervelocity impacts (say, radius and depth of craters, the range of crater ejecta and thickness of ejecta deposits) and input parameters such as impact velocity, characteristics of impactors, material strength, gravity, and others. Experimental studies of impacts, analytical solutions and code calculations of the relationships between mass, momentum, and energy, as well as the dimensional analysis of the variables, are the primary tools to derive scaling theories. In such a way the scaling laws provide a bridge between laboratory experiments and large planetary impact events, which are impossible to reproduce in actual practice.

We consider an impact occurring in a gravity field of magnitude $g$. When the strength of the target material predominates over gravitational effects, the latter plays a negligible role in the process of crater formation. Otherwise the gravitational forces regulate the cratering process. These two extremes are often referred as "strength-dominated and gravity-dominated regimes" depending on the scale of the impact event. In our case, the diameters of most projectiles vary from ten centimeters to some meters, so that craters on asteroids will form in the strength regime at small sizes and tend to transition to the gravity regime at larger crater sizes. We use simple scaling relationship between the crater and projectile diameter taken from Richardson et al. (2005):

$D_{cr} = 30 D_p.$  (5)

Judging from Figure 16 in the referred work, this relation corresponds to transient conditions between pure strength and pure gravity regimes for various types of target materials from loose sand to competent rocks and well satisfies the numerical hydrocode simulation of the cratering events on asteroids.

Now using Eq. (5) we can estimate for the given asteroid the total area of craters $S_{tcr}$ formed in the time $T_{OC \to S}$:

$$S_{tcr} = 225\pi \sum D_p^2 N_p,$$

where $N_p$ is the number of impactors with diameter $D_p$. As $N_p$ varies directly with $D_a^2$, it follows that $S_{tcr}$ and the total area of the asteroid surface, $S_a$, are directly proportional. For the target asteroids of 20–500 km in size, the ratio of $S_{tcr}/S_a$ is approximately the same and equals ~0.004. We verified this inference using other power-law scaling approximations that link crater diameters, impact velocity, and gravitational acceleration (Eq. 16 in Housen et al., 1979; Eq. 22a in Holsapple, 1993) and obtained similar results: the craters forming in $T_{OC \to S} \sim 1.5 \times 10^5$ yr occupy less than 1% of the asteroid surface even if they do not overlap. That is, the impact flux has not yet had time to destroy the original surface of asteroids completely. Constancy of $S_{tcr}/S_a$ ratio for asteroids of various sizes reflects the fact that the small-crater population dominates among the craters in question.

A portion of the kinetic energy of the impactor (about 10 %) is spent on ejection of material during crater formation (O'Keefe and Ahrens, 1977). Dimensional analysis leads to the following dependence of the velocity, $v$, of excavated material when it passes through the original target surface at a distance, $x$, from the impact point (Housen et al., 1983; Housen and Holsapple, 2011):

$$\frac{v}{U} = \left[ \frac{x}{R_p} \left( \frac{\rho}{\delta} \right)^\nu \right]^{-1/\mu} f(x/R_{cr}), \tag{6}$$

where $U$ is the impact velocity, $R_p$ is the impactor radius, $\rho$ and $\delta$ are the mass densities of the target and impactor materials, respectively, $R_{cr}$ is the crater radius, and $f(\cdot)$ is some function that is not determined by means of dimensional analysis. Housen and Holsapple (2011) stress that Eq. (6) describes the ejection velocity as a function of a launch position $x$ both for the strength and for the gravity regimes, as the crater radius $R_{cr}$ depends on the strength and gravity effects. The authors note also that the power-law approximation with exponent $-1/\mu$ does not work near the crater edge where the ejection velocity goes to zero. It is therefore assumed that the function $f(\cdot)$ in Eq. (6) can be expressed as $(1-x/n_2 R_{cr})^p$. That is, Eq. (6) takes the form (Housen and Holsapple, 2011):

$$\frac{v}{U} = C_1 \left[ \frac{x}{R_p} \left( \frac{\rho}{\delta} \right)^\nu \right]^{-1/\mu} \left( 1 - \frac{x}{n_2 R_{cr}} \right)^p, \tag{7}$$

where constant $C_1$ and exponents $\nu, \mu, p$ should be found from impact experiments, $n_2$ takes separate values $n_{2,S}$ and $n_{2,G}$ for the strength and gravity regimes, $x$ varies from $n_1 R_p$ to $n_2 R_{cr}$ and $n_1$ is assumed to be approximately 1.2.

The scaling law for the mass of ejected material having launch position less than $x$ is $M(<x) = k\rho x^3$ (Housen and Holsapple, 2011). But $M(x)$ goes to zero if $x \to n_1 R_p$ because material moves in a downward direction inside of $x = n_1 R_p$ and is not ejected. With this in mind the authors modify this expression to the form:

$$M(<x) = k\rho \left( x^3 - [n_1 R_p]^3 \right), \tag{8}$$

where $k$ is a constant that should be estimated from experiments. The authors point out that $M(<x) = M(>v)$, that is, the ejected material has velocity greater than arbitrary value, $v$, because the ejection velocity decreases monotonically with increasing $x$ according to Eq. (7).

In this scaling theory the total mass of ejecta and the crater radius are coupled variables:

$$M_{cr} = k_{cr} \rho R_{cr}^3, \tag{9}$$

where, $k_{cr}$ is an empirical constant less than 1 (see the referred work for more details). The scaled mass of ejecta which is launched faster than the corresponding velocity is

$$\frac{M(>v)}{M_{cr}} = \frac{k}{k_{cr}} \left( \frac{x^3}{R_{cr}^3} - \frac{[n_1 R_p]^3}{R_{cr}^3} \right). \tag{10}$$

Owing to the simple relation between the crater and projectile diameter (Eq. 5), which we use to characterize the impact events on asteroids, the ejection velocity and the scaled mass of ejecta depend neither on the size of crater nor the impactor size. Since $R_p = \frac{1}{30} R_{cr}$ and the launch position $x$ is measured in units of crater radius (i.e., $x = qR_{cr}$), then Eqs. (7) and (10) are reduced to

$$\frac{v}{U} = C_1 \left[ 30q \left( \frac{\rho}{\delta} \right)^\nu \right]^{-1/\mu} \left( 1 - \frac{q}{n_2} \right)^p, \tag{11}$$

$$\frac{M(>v)}{M_{cr}} = \frac{k}{k_{cr}} \left( q^3 - [n_1/30]^3 \right), \qquad n_1/30 \leq q \leq n_2. \tag{12}$$

Distribution of the scaled mass of material which is ejected faster then the given velocity $v$ is shown in Figure 5. The calculations were performed with the following assumptions. For the sake of simplicity we suppose $\rho = \delta$. The impact velocity $U$ is equal to 5 km/s, average relative velocity in the asteroid belt. The numerical values of the empirical parameters were chosen according to the data listed in Table 1 from Housen and Holsapple (2011): $C_1 = 0.5$, $\mu = 0.4$ and $p = 0.3$ – they are all for porous, regolith-like materials; $k/k_{cr} = 0.5$ is reasonable for common soils and could be the upper limit for asteroid porous regolith; $n_1 = 1.2$ and $n_2 = 1$. Figure 5 illustrates the important property of the ejecta model developed by Housen and Holsapple (2011): at least for small impact events when the simple "cube-root" scaling law (Eq. 5) works, almost all the mass of the impact crater ejecta accumulates on the target asteroids with diameter larger than 10 km. Only a small fraction of material launched from near the crater center moves faster than the escape velocity $v_{es}$, and can leave the asteroids. The substantial regolith layer seen covering small asteroids such as Gaspra, Eros, and Ida is a good corroboration of this conclusion (Sullivan et al., 2002).

If the launch position varies in the range $x_1 \leq x \leq R_{cr}$, the volume, $\Delta V$, of material ejected from the given crater is

$$\Delta V = \frac{1}{\rho}\Delta M = \frac{1}{\rho}[M(<R_{cr}) - M(<x_1)] = kR_{cr}^3(1 - q_1^3), \qquad (13)$$

where $q_1$ corresponds to the ejecta velocity $v(q_1) \sim 0.9v_{es}$ and $k = 0.3$ in accordance with the data from Table 1 in Housen and Holsapple (2011). The total volume of material produced by impacts into the given asteroid is

$$\Delta V_t = \sum \Delta V N_p = 15.8(1 - q_1^3)\sum D_p^3 N_p. \qquad (14)$$

For simplicity we assume that after a lapse of a sufficiently long time period the ejecta blankets cover asteroid surfaces more or less uniformly. Then the average thickness of this layer on the asteroid with diameter $D_a$ is approximately

$$H_b \approx \frac{\Delta V_t}{\pi D_a^2}. \qquad (15)$$

We plotted Figure 6 using Eqs (11–15) and the same numeric values of the foregoing empirical parameters. Shown in Figure 6a is the dependence of total volume, $\Delta V_t$, of ejecta on the time interval $\tau$ between impacts. As is seen from this Figure and Figure 4c, just small impactors provide the main growth in $\Delta V_t$ due to the higher frequency of collisions with asteroids; the larger the cross-section of the asteroid, the larger the total volume of ejected material. The average thickness of the ejecta blanket as a function of the asteroid diameter is shown in Figure 6b. Thickness of the blankets which are accumulated on the asteroids of various sizes in the time span $T_{OC \to S} \sim 1.5 \times 10^5$ yr averages from $\sim 0.3$ cm for the 20-km asteroid to $\sim 0.5$ cm for the 500-km asteroid.

To verify the correctness of our approach we compared our calculation of the thickness of asteroid regolith formed over an interval of $\sim 10^9$ yr on a 20-km body (see Figure 6c) with analogous computation of Richardson (2011). If we take into account the differences in baseline assumptions, the calculation algorithms, and the used numerical values of the empirical parameters, one can conclude that the theoretical curves shown in Figure 6c are in satisfactory agreement (i.e., both indicate accumulations of a meters to tens of meters thick regolith). Indeed, we can also deduce, as Richardson (2011) has done, that the regolith layer is generated on decimeter, meter, and decameter scales after a lapse of ~ $10^7$, $10^8$, and $10^9$ yr, respectively, on asteroids of 20 km in size. Extrapolating to shorter time intervals we derive a subcentimeter scale for regolith thickness on timescale of ~ $10^5$ yr. The result for asteroids of various sizes is represented in Figure 6b.

So, over the lifetime of asteroids, fresh subsurface material is ejected and accumulating on their surfaces due to the ongoing cratering process. On asteroids of various sizes, the thickness of this layer formed in the time of $T_{OC \to S} = 1.5 \times 10^5$ yr amounts to some thousands of micrometers. That is, the original surface, which should have been irradiated by the solar wind, is episodically covered by layers of fresh material. Penetration of visible light into silicate minerals is on the order of a few tens of micrometers, say, ~ 50 μm for wavelength of 0.5 μm. This topmost layer of asteroid regolith is formed in ~ $5 \times 10^3$ yr on asteroids with $D_a$ ~ 20 km (and even faster on larger asteroids). This time is much shorter than the $T_{OC \to S} = 1.5 \times 10^5$ yr, therefore this topmost optically active layer, as with the underlying layers, is actually immature material.

On the other hand, the layer of ejecta deposited on asteroids in the time $T_{OC \to S}$ is sufficiently thin on the body scale. Since crater formation is a stochastic process, sediment thickness can vary over asteroid surface, somewhere there may be more, somewhere less than the average. Because of natural topography, an asteroid's surface area is larger than the area of an equivalent sphere; therefore some asteroid terrains could be free from ejecta deposits. A detailed description of regolith formation on asteroids is beyond the scope of this work (see the excellent works of Richardson et al. (2005; 2007) and Richardson (2009)). Nevertheless we can infer that on the same weathering timescale, the maturation level of an asteroid as a whole will be less than that of an undisturbed and unshielded surface exposed to solar wind.

Such an asteroid, in the case of disk-unresolved observations, would look like a slightly weathered body. Of course, we can try to increase the maturation level of an asteroid surface by expanding the weathering timescale, say, up to $\sim 8\times 10^8$ yr (this value from Paolicchi et al., 2007). However, over this time the original surface will be destroyed by ensuing impacts, the impact ejecta blanket thickness will run up to some meters, a new reference surface will be formed, and… one can start reading this Subsection from the beginning.

The formation of regolith on asteroids is accompanied not only by the ballistic motion of particles, but being already on the asteroid surface, the particles are involved in other types of movements such as downslope movement, shaking, mixing, sorting by particle size - all triggered by impact-induced seismic events.

*3.3. Seismic activity of asteroids and regolith motions*

A small fraction of the kinetic energy of an impactor (~ 0.1% or even less) is converted into seismic energy that in the form of seismic waves travels throughout an asteroid. Depending on projectile and asteroid masses, the impacts generate the seismic vibrations of asteroid surface in the regional or global scale. Recently, Richardson et al. (2005; 2009) have developed the theory of seismic processes following impacts into asteroid-sized bodies. Asteroids are believed to be fractured bodies having internal structure similar to upper lunar crust: a relatively thin regolith layer covers a thick megaregolith formation, which gradually turns into fractured bedrock. Each impact into an asteroid is the source of the body spherical waves, which penetrate through the asteroid interior and undergo multiple scattering in the randomly inhomogeneous medium, and refraction and reflection at the internal interfaces and asteroid surface. Owing to the small sizes of asteroids and to the fact that the body wave velocity is about 2 – 3 km/s in fractured medium, the asteroid interior is quickly filled up with a low-frequency "seismic sound". Most asteroids are likely extremely dry as is the lunar interior, therefore the seismic energy dissipation is low and the asteroid is capable of retaining the impact excitation longer. In the case of asteroid Eros, the duration of synthetic seismograms obtained away from an impact site varies from a few minutes to about an hour depending on the impactor size (Richardson et al., 2005). Reverberations of the lunar surface that have been observed during lunar seismic experiments could last longer than an hour and half (Toksöz, 1975; Dainty et al., 1974).

When the body waves reach the surface, they generate cylindrical seismic waves that travel along an asteroid's surface within an upper shell having thickness of the order of the wavelength of surface oscillations. The surface waves are characterized, in comparison with the body waves, by a low velocity, high intensity, low-frequency spectrum, a rapid attenuation with depth, and long-term oscillations. At large distances from the impact site (or epicenter in the given context), surface waves become much intense than body waves and can cause more extensive modification of natural topography. Once originated, a surface wave diverges from the epicenter in all directions and, rounding the entire asteroid, may meet at the antipodal point where a new seismic event of lesser intensity occurs. Figure 7 illustrates the effect of the multiple passages of a surface harmonic wave between the epicenter and its antipode on spherical bodies with the same fractured structure as described above. In our illustrative example, the phase velocity of the wave and its period are 0.49 km/s and 5 s, respectively. It is supposed the wave penetrates through a homogeneous medium, so that the wave velocity does not depend on distance from the epicenter. The maxima of the curves shown in the Figure decrease exponentially due to dissipation and scattering of seismic energy. The seismic attenuation parameter, $Q = 2000$, and the seismic diffusivity, $K_s = 3$ m$^2$ s$^{-1}$, were taken from Richardson et al. (2005). This example demonstrates that a single impact event (specially a large event) is able to induce regolith particle motions, even on the antipodal side.

Let us consider a particle motion in a surface wave of Rayleigh type. There are other types of surface waves known from terrestrial seismology, but they need more specific environments than those for Rayleigh waves to be generated (Sheriff and Geldart, 1982). If Rayleigh wave propagates through a semi-infinite elastic and isotropic layer, then each unit of the medium describes a closed elliptic trajectory in a plane normal to the surface and parallel to the traveling direction. The motion of the units is retrograde, ellipses have maximal size at the surface, and the major semi-axis of the ellipses is also normal to the surface. However, the physical properties of natural solids or structures – such as density and stiffness to tensile stress – vary with depth and distance, so that Rayleigh waves become dispersed and, in particular, the surface particles experience more complicated motion than that in ellipses (see Figure 8). Since fractured asteroids are certainly non-isotropic bodies, the loosely coupled particles of asteroid regolith will most likely move irregularly under the action of Rayleigh waves, approximately in such a way as shown in Figure 8. If the amplitude of Rayleigh waves is larger than the size of particles, the initial position of a particle and its final position, occurring when the wave passes, will be different (as shown in Figure 8), and so this wave process tends to mix adjacent particles. Since particles shift most in the vertical plane (in relation to the surface plane) the fresh subjacent particles tend to mix with the topmost ones.

The long-term seismic vibrations of asteroid surfaces produce downslope motion of regolith resting before impact on inclined planes (Richardson et al., 2005 and references therein). As a result, fresh unweathered material outcrops due to this process. Richardson et al. (2005) note that this seismically triggered downslope motion is a combination of horizontal sliding and vertical hopping of the upper regolith layer. In other words, under the specified amplitude and frequency of seismic vibrations, a perturbing force predominates over a cohesion force and regolith particles gain one more degree of freedom being revealed as sliding and hopping.

Under the seismic vibrations, the last type of regolith motion is obviously the prevalent event and occurs even if a surface slope angle is zero. One can expect that hopping (or shaking) of regolith particles results in their intermixing and size sorting if they differ in size and/or density (e.g., the reviews of Ottino and Khakhar, 2000 and Kudrolli, 2004 and references therein; Murdoch et al., 2011; Richardson et al., 2011). The larger particles under vertical vibration of granular medium tend to accumulate at the top of the surface; this case is called the Brazil nut effect (e.g. Rosato et al., 1987). However, the reverse Brazil-nut effect may also occur. If the density of large particles is much greater than that of the small ones and the amplitude of vibration is greater than the size of large particles, they begin to move downwards while the small particles rise to the surface (Schröter et al., 2006; Breu et al., 2003). Features consistent with regolith migration and segregation on the scale of centimeter to several tens meter have been documented on asteroids Eros and Itokawa (Asphaug et al., 2001; Miyamoto et al., 2007). Richardson et al. (2005) have demonstrated the efficacy of seismic erosion and eventual erasing of small impact craters on Eros-sized asteroids. Moreover, the latter is apparently valid for the sub-kilometer-sized asteroid Itokawa in spite of its low density ( ~ 1.9 g/cm3), high porosity ( ~ 40%) and a very likely rubble-pile structure (Abe et al., 2006; Fudjivara et al., 2006). One might expect that this type of asteroid internal structure substantially attenuates seismic energy. Nevertheless, the lack of small craters on Itokawa is consistent with their erasure by seismic shaking due to, in particular, the cumulative effects of many small impacts (Michel et al., 2009).

The seismically induced motion of regolith particles becomes significant only if a dimensionless seismic acceleration ratio $a_s / g > 1$ ($g$ denotes acceleration due to gravity of the given target asteroid). Richardson et al. (2005) have estimated the minimum impactor size, $D_{p,\min} \propto D_a^{5/3}$, necessary to achieve global seismic acceleration equal to $g$ on asteroids of various sizes without their total disruption. Thereby impactors with diameter larger than $D_{p,\min}$ are able to cause global seismic effects throughout the volume of an asteroid, destabilizing all slopes on the surface. It is obvious that each such event greatly reduces space weathering effect for the asteroid surface as a whole due to the extensive surface modification. During the relatively short time span $T_{OC \to S}$, global shaking of surfaces occurs about ~ 1000 and 100 times on asteroids of 20 and 30 km in diameter but is an almost unlikely event if $D_a$ > 40 km. This happens because the vast majority of projectiles falling on the asteroids over this time period are small, and their surface-modifying seismic effects are local by their very nature. It makes sense therefore to estimate the area around impact sites where seismic acceleration is equal to or more than *g*. For this we use the following equation between the seismic energy, $E_s$, and the distance, *r*, from the impact site (Yanovskaya, 2008):

$$E_s = \eta E_k = \frac{8\pi^3 A^2}{T} \rho v_p r^2 \exp(2\alpha r). \tag{16}$$

This equation is written for the first highest-power body wave having the amplitude *A* and period *T* and propagating with velocity $v_p$. Other parameters of Eq. (16) are the following: $E_k$ is the kinetic energy of an impactor, $\eta$ is the impact seismic efficiency factor, $\rho$ is the mass density of the target asteroid, and $\alpha$ is the attenuation coefficient of the body wave. Keeping in mind that the maximum seismic acceleration, $a_s$, can be expressed in terms of the maximum displacement *A* of the medium and frequency *f*=1/*T*, that is, $a_s = 4\pi^2 f^2 A$, we can rewrite Eq. (16) as follows:

$$r \exp(\alpha r) = a_s^{-1} \left( \frac{2\pi f^3 \eta E_k}{\rho v_p} \right)^{1/2}. \tag{17}$$

If we set $a_s$ to be equal to *g* in Eq. (17), then *r* will equal $r_g$, the "seismic radius" around the impact site where the seismic and gravity accelerations are equal. Clearly, if *r* < $r_g$, then $a_s$ > *g*. Note that *r* (or $r_g$) is, in fact, the straight-line chord distance between the impact site and a point on asteroid surface (the asteroid body is supposed to be spherical). We believe also that the range of elastic deformations starts at a distance of $2R_{cr}$ from the point of impact, so that the seismic area around the impact site where the surface modification processes could occur, is $S_g = \pi(r_g^2 - D_{cr}^2)$.

It is interesting to estimate the total seismic area, $S_{tg}$, around all impact craters originated on the given asteroid in the time $T_{OC \to S}$. This value scaled to the total asteroid area is

$$S_{tg}/S_a = \frac{1}{D_a^2} \sum_p (r_g^2 - D_{cr}^2) N_p, \tag{18}$$

where $N_p$ is given by Eq. (2). The numerical values of the parameters required to calculate this ratio were taken from Richardson et al. (2005) and Richardson (2009). Namely, the mean mass density of asteroids and projectiles, ρ and δ, are assumed to be the same and are equal to 3 g/cm$^3$; the impact velocity of $v$ = 5 km/s; the body wave velocity of $v_p$ = 0.5 km/s, i.e., the typical value for the $P$ wave in loose material; η = 10$^{-4}$ and α = 0.069 km$^{-1}$. Synthetic seismograms simulating the surface vibrations of fractured asteroid body have a frequency between 1 and 100 Hz with a peak near 10 – 20 Hz (Richardson et al., 2005). So we used in the calculations three values $f$ = 10, 15, and 20 Hz.

Shown in Figure 9 is dependence of the $S_{tg}/S_a$ ratio on asteroid diameters at various values of the surface vibration frequency. As is seen from the Figure, $S_{tg}/S_a > 1$ for asteroids with diameters from 20 to several hundred kilometers. This suggests that during the time $T_{OC \to S} \sim 1.5 \times 10^5$ yr every surface element of an asteroid experiences multiple impact excitations with sufficient seismic magnitude to generate the various types of regolith motions discussed above. Specifically, regular mixing of the exposed and subsurface fresh regolith particles appreciably reduces the optical effect of space weathering of the main belt asteroids.

Let us return to Figure 4b in order to consider a correlation between collision resurfacing and surface maturation from another viewpoint. For the same time $T_{OC \to S}$ the 500-km body in the lunar orbit experiences fewer collisions by approximately two orders of magnitude than in the main belt. Thereby if the thickness of ejecta blankets on the body averages ~ 0.5 cm at the distances of the asteroid belt (see Figure 6b), then the value will be less by a factor of 100 at the distance of 1 AU, that is, only 0.005 cm. Since the impact velocity is about 25 km/s at 1 AU, the seismic radii (Eq. 17) are about five times more here than those in the asteroid belt. However, the total seismic area $S_{tg}$ on the 500-km body is nevertheless less by a factor of $\sim 4$ as against the same value in the asteroid belt and, consequently, $S_{tg}/S_a < 1$ for the body in the lunar orbit. The latter means that the seismic events occur in the immediate vicinity of the impact craters and $\sim 75\%$ of the body surface can be accepted as undisturbed. Keeping in mind the trace amount of material deposited for the time $T_{OC \to S}$, one can conclude that the 500-km body at 1 AU maturates according to the lunar type. This conclusion is valid, of course, for asteroids of any size provided that they are located in the lunar orbit.

## 4. Discussion and conclusion

If we make a supposition that the spectra of S asteroids are the modified spectra of ordinary chondrites due to space weathering effects, then our analysis allows the following paradox to be clarified: Why does the spectral modification process of main belt asteroids apparently terminates with S-type spectra on a timescale of $\sim 10^5$ yr passes and does not proceed to "completion"; i.e., lunar-type spectra over the next $\sim 10^6$ yr and beyond? This is especially puzzling since both these time intervals are much shorter than the collisional lifetime ($\sim 10^9$ yr) of asteroids larger than about 10 km in size (O'Brien et al., 2005).

As we have demonstrated, the difference in the degree of optical maturation of the Moon and asteroids results from the unique properties of their regoliths. The uppermost layer of the lunar surface survives for $\sim 10^7$ yr before burial by another layer, so that lunar regolith particles have time to accumulate detectable amount of reduced iron and proceed to lunar-type maturation (Borg et al., 1976). Unlike the Moon, asteroid regoliths are characterized by higher particle mobility due to the high frequency of non-catastrophic collisions with projectiles and low gravity of asteroid targets (Richardson et al., 2005).

We then asked ourselves a question: What is happening on asteroid surfaces on the space weathering timescale of ~ $10^5$ yr, i.e., in the time needed to make any "chondritic" spectrum similar to the S-type? The expected time of impact onto asteroids turns out to vary from some hundred days to about ten hours depending on their size. Therefore asteroid surfaces cannot be considered as undisturbed on this weathering timescale. We have demonstrated above that impact activated motions of regolith particles hamper the progress of optical maturation of the asteroid surfaces. Apparently, the fact that the space weathering timescale of ~ $2\times10^9$ yr inferred from S-asteroid spectral observations (Willman and Jedicke, 2011) is much longer than the laboratory estimation $T_{OC\rightarrow S}$ based on ion irradiation experiments (Hapke, 1973) may be a direct consequence of regular rejuvenation of asteroid surfaces. Admittedly, in our opinion, the astrophysical timescale is self-contradictory. For this very long time interval, the original (reference) surface of asteroids has been destroyed during the first $10^7$ yr, the thickness of the regolith will reach several tens of meters, and global resurfacing events will iteratively happen in the history of asteroids with diameters up to 70 – 100 km (Richardson et al., 2005; Richardson, 2011). The maturation level of asteroid surfaces may be a compromise resulting from a competition between impact resurfacing and solar wind darkening, and after reaching some steady state after a relatively short time (say, several times × $T_{OC\rightarrow S}$), no longer depends on time.

One can expect the "weathered skin" does not cover the asteroid surfaces with the same degree of homogeneity as on the Moon given the high mobility of asteroid regoliths. Thus, variations of disk-integrated spectral characteristics among S asteroids appear to be caused by variations of surface material composition, particle sizes and porosity of the topmost regolith layer, perhaps vitrification, and agglomeration of regolith particles rather than optical maturation of their surfaces.

As is seen from Figures 2 and 3 the *b* parameter is a measure of the maturation level of a surface. If *b* is sufficiently low, then $\tau_{Fe}$ can be much less than $\tau_h$ and spectral alteration occurring due to space weathering will be less (or, at least, comparable with) the natural spectral variations of mineral constituents forming the asteroid's surface. This problem becomes relevant when modeling asteroid spectra and not only for the S-types. For example, V asteroids demonstrate low maturation levels, as seen in the best matches between the observed and theoretical spectra (Shestopalov et al., 2008). In the case of S asteroids, the *b* parameter appears to be less than 0.002, i.e., the value used for undisturbed ordinary chondritic surface (see Figure 2). The numeric simulations associated with the modern physical theories (Shkuratov et al., 1999; Hapke, 2001) are needed to estimate maturation level more accurately for each S-type asteroid studied.

The traits of optical maturation on several asteroids have been found from multi-color images and spatially resolved spectrometry obtained by spacecrafts (Chapman, 2004; Clark et al., 2003; Ishiguro et al., 2007). The spectral contrasts of Eros' dark terrains are believed to be determined by not only lunar-like optical maturation, but also a dark spectrally neutral constituent, presumably shock-modified troilite (Clark et al., 2001; Bell et al., 2002). In turn, spectral analysis of Eros' areas leads to an ordinary chondritic composition (McFadden et al., 2001; Bell et al., 2002), though the freshest bright material shows significant spectral differences from the proposed meteorite analogues (Clark et al., 2001). Shestopalov (2002) has interpreted a strong correlation between albedo (at 760 nm) and 950-nm/760-nm reflectance ratio previously reported by Murchie et al. (2002) for spatially resolved areas in the northern hemisphere of asteroid Eros. This "Eros line" on the albedo – color plot arises from the following fact: the larger the regolith particles on the asteroid surface, the higher the cumulative amount of reduced iron in these particles. The result quantitatively shows the difference in maturation of the asteroid and lunar regoliths. For the lunar soil, the finest fraction is enriched in reduced iron relative to the large size fraction and dominates optical properties of the lunar bulk material (Hapke, 2001; Noble et al., 2001) The "unusual" behavior of the Eros' regolith particles may be a response to seismic vibrations of the asteroid surface (the Brazil nut effect): the larger the particle, the likelier it is to survive on the surface; the longer its exposure time, the greater the cumulative amount of $Fe^0$ in the particle.

Unlike asteroid Eros, the surface of Itokawa, the sub-kilometer-sized near-Earth asteroid, shows extensive areas of coarse materials, from centimeter-sized pebbles to boulders of a several tens of meters in size (Miyamoto et al., 2007). Mainly due to this, the Itokawa surface is less mature then that of Ida, Gaspra and, probably, Eros since optical maturation of solids is less effective in comparison with fine-grained regolith (Hapke, 2001; Ishiguro et al., 2007). Maturation level correlates with the surface morphology (Sasaki et al., 2007; Ishiguro et al., 2007) which in turn is a result of various types of granular motions (as landslide-like migrations, particle sorting and segregation and others) triggered mainly by impact-induced seismic vibrations (Miyamoto et al., 2007). All these undoubtedly significant nuances obvious "at proximate examination" play a minor role in the determination of Itokawa's composition inferred from ground-based spectral observations. The inference that the rocks of Itokawa in whole are similar to LL chondrites (Binzel et al., 2001; Abell et al., 2007) was made well before surface specimens were returned by the Hayabusa probe (Nakamura et al., 2011). Moreover, the telescopic investigations of the asteroid (Abell et al., 2007) reveal complex thermal evolution of a parent body before its disruption, one of the pieces of which is known today as Itokawa. Returned samples support the idea of complex regolith evolution, as individual grains have compositions ranging from petrologic grade 3 to 6 and show variable amounts of nanophase iron from grain to grain (Nakamura et al., 2011).

Viewing the asteroid literature one can see that spectroscopic analysis fairly states whether a given S asteroid has chondritic composition (e.g., Binzel et al., 2009) or does not (e.g., Hardersen et al., 2006), regardless of whether the investigators take into account optical effect of space weathering or not. It is not surprising since the maturation level is so low that maturation does not appreciably shift band center positions in asteroid spectra; essentially complete vitrification of pre-existing minerals is required for significant alteration of these spectral parameters (Gaffey, 2010). Asteroid space weathering also does not appreciably attenuate the intensity of faint absorption bands in the visible region (Cloutis et al., 2011 and reference therein; Shestopalov et al., 1991; Hiroi et al., 1996).

The large S asteroids of the main belt reveal a great diversity of compositions, and the probable parent bodies of ordinary chondrites are only one of the subtypes of this asteroid population (Gaffey et al., 1993). It is appropriate to mention here that the direct numerical experiment, namely simulation of the space weathering optical effect without any suppositions about the character of asteroid regolith evolution, demonstrates that reflectance spectra of ordinary chondrites cannot be converted into the spectra of large S asteroids (Shestopalov and Golubeva, 2004). Moreover, S asteroids show compositional trends along the perihelion and mean distances from the Sun (Golubeva and Shestopalov, 1992) whereas possible meteorite sources appear to be also located in the outer part (~2.8 – 3.0 AU) of the asteroid belt (McFadden et al., 1985; Golubeva and Shestopalov, 1992; 2002). Recently Nesvorný et al. (2009) have shown that shocked L chondrite meteorites, the fragments of a disrupted S asteroid, could be transformed to the Earth-crossing orbits via 5:2 resonance with Jupiter at 2.823 AU.

Unlike our simplifying assumptions (though in the calculations we used the conservative values of problem parameters and did not oversimplify the significance of these and other factors), asteroids differ in composition, internal structure, mechanical and seismic properties. The surface of each asteroid is, in fact, a snapshot of its collisional and proton irradiation histories and the surface maturation level is in its own way a product of the confluence of these processes. However, there is no need to simulate simultaneously these processes to quantify the maturation level of any asteroid surface. In a practical sense it is much easier to evaluate this surface attribute by modeling asteroid spectra using scattering theories (Shkuratov et al., 1999; Hapke, 2001). Such an investigation could characterize the surface maturation of asteroids of various compositions and belonging to different age groups.

Our results suggest that space weathering processes cannot give rise to color diversity in the main asteroid belt, however we can face the opposite situation in the Edgeworth–Kuiper belt. Ion irradiation experiments simulating solar wind effects on complex carbonaceous compounds and hydrocarbon ices (i.e., potential constituents of the trans-Neptunian objects' and Centaurs' surface materials) also show optical maturation trends, though differing from those for silicate minerals (Moroz et al., 2004). Notably, the probable weathering timescale estimated is short, $\sim 3 \times 10^5$ yr at $\sim 40$ AU (Moroz et al., 2003). Keeping in mind that the impact velocity here is $\sim 0.5$ km/s and the collision rate may be less than in the asteroid belt (Davis and Farinella, 1997), space weathering might produce diversity of the optical properties of the Edgeworth–Kuiper belt bodies.

**Acknowledgments**

We express our thanks to anonymous reviewer for comments and interesting discussion.

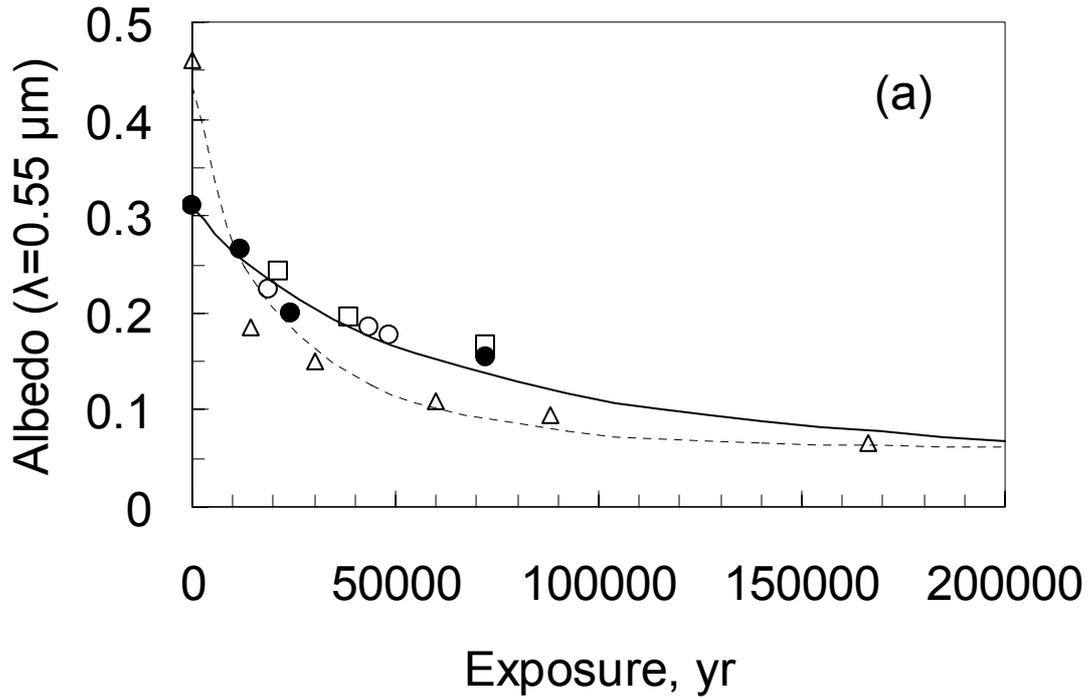

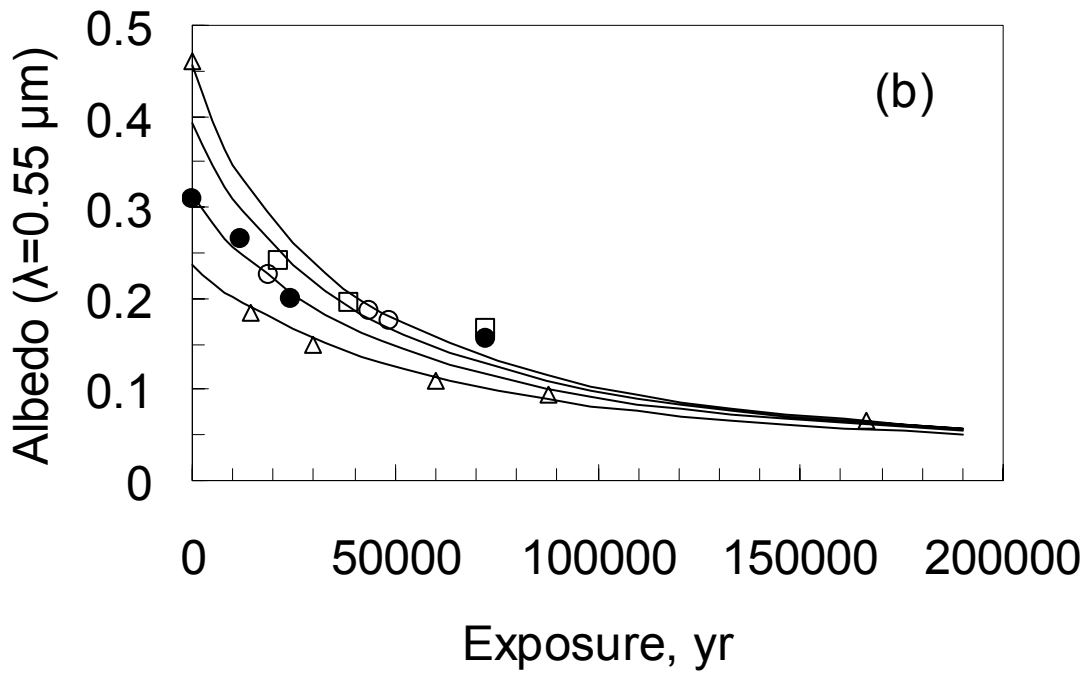

Fig. 1. Albedo of olivine basalt powder versus the exposure time for varying H-ion beam densities and particle sizes in simulations of the darkening of silicate powders by solar wind irradiation at a heliocentric distance equal to 1 AU. These original data were adopted from Hapke (1973): filled circles denote ion beam density equal to 0.33 mA cm$^{-2}$ and particle size less than 37μm; open circles: 0.17 mA cm$^{-2}$, < 37μm; squares: 0.09 mA cm$^{-2}$, < 37μm; triangles: 0.50 mA cm$^{-2}$, < 7μm; (a) Dashed line fits the fine-grained samples at $b_{max} = 0.08$ μm and $T=2\times10^5$ yr whereas solid line fits the coarse-grained samples at $b_{max} = 0.032$ μm and $T=2\times10^5$ yr; (b) Numerical calculations for various albedos of host material as a function of the exposure time were carried out at $b_{max} = 0.038$ μm and $T=2\times10^5$ yr.

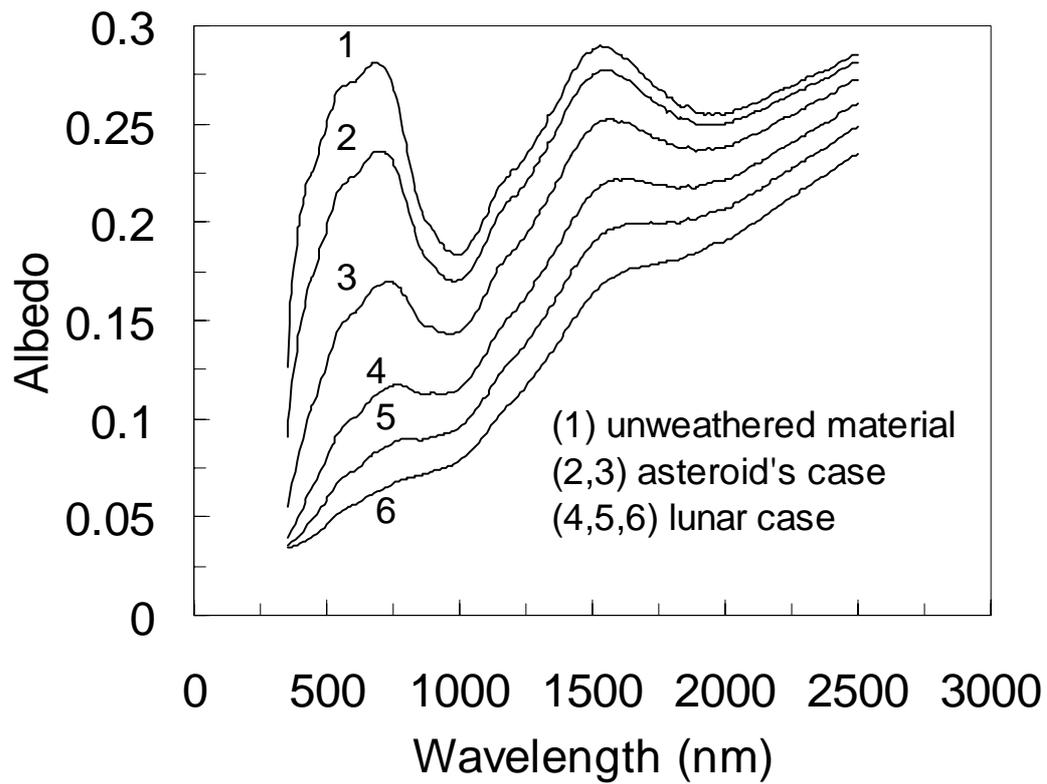

Fig.2. Transformation of reflectance spectra of the Vavilovka LL6 ordinary chondrite meteorite with increasing concentration of fine-grained metallic iron within the rim of host particles; (1) is the original spectrum of Vavilovka (Gaffey, 1976); spectra (2 – 6) were calculated using the original spectrum (1) and $b$ = 0.00247, 0.00836, 0.01748, 0.0266, and 0.038 μm.

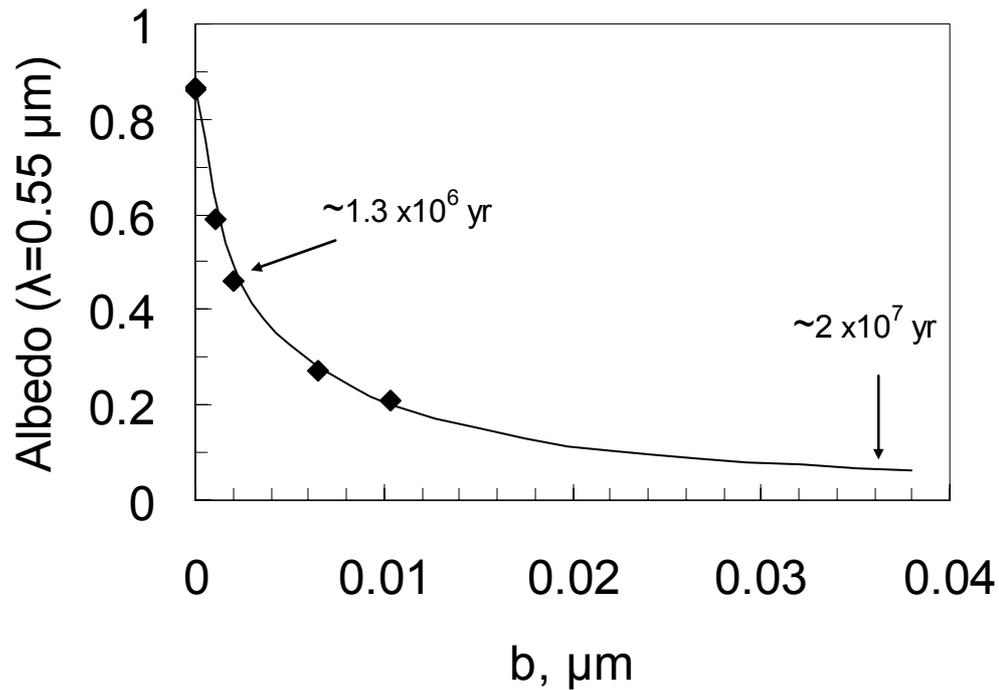

Fig. 3. Albedo of olivine pellet samples before and after pulse laser irradiation in a simulation of space weathering effect by micro-particle impacts (Sasaki et al., 2001) versus parameter $b$ which has been estimated by Shestopalov and Sasaki (2003). The five albedos (diamonds) correspond to (from top to bottom): non-irradiated sample, 15-mJ irradiation, 30-mJ irradiation, five and ten repetitions of 30-mJ irradiation treatment. Numerical calculation (line, this work) was carried out at $b_{max} = 0.037$ μm and $T=1.9 \cdot 10^7$ years.

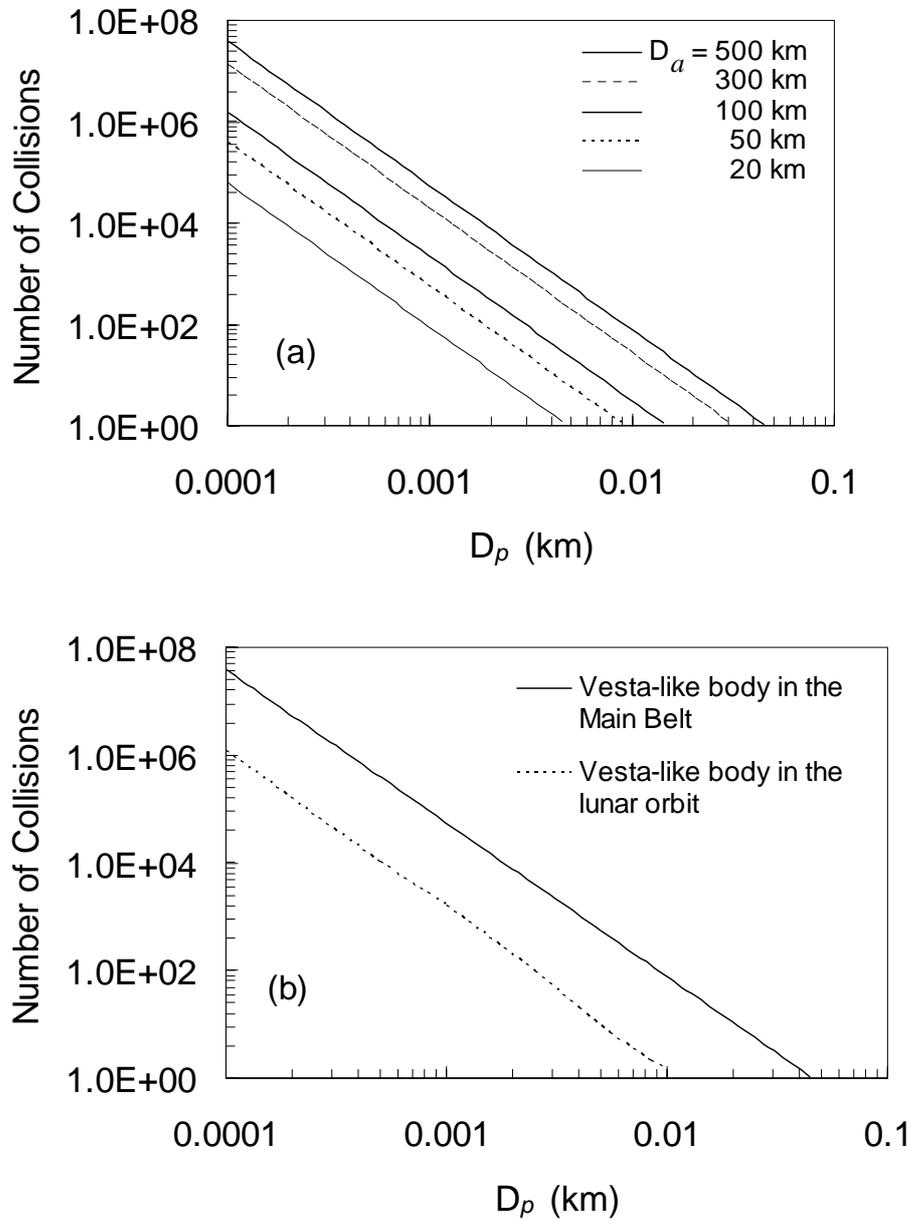

Fig. 4. (a) The number of collisions between asteroids of given diameter ($D_a$) and impactors of diameter ($D_p$) or larger in time interval $T_{OC \to S} = 1.5 \; 10^5$ years; (b) The number of collisions between 500-km body and impactors in the asteroid belt and in the lunar orbit during the same time interval. The dotted line is linear extrapolation based on the NEA size-frequency distribution taken from Werner et al. (2001);

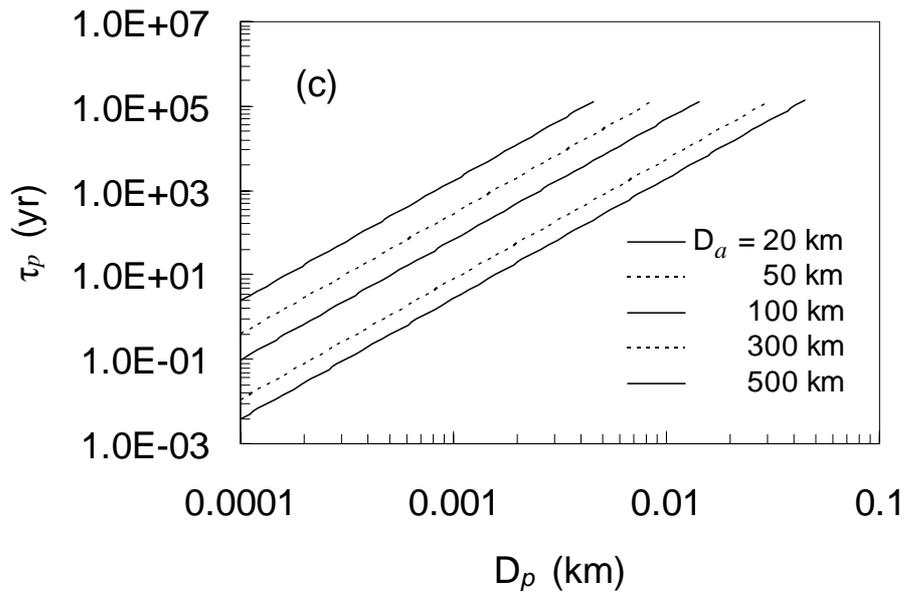

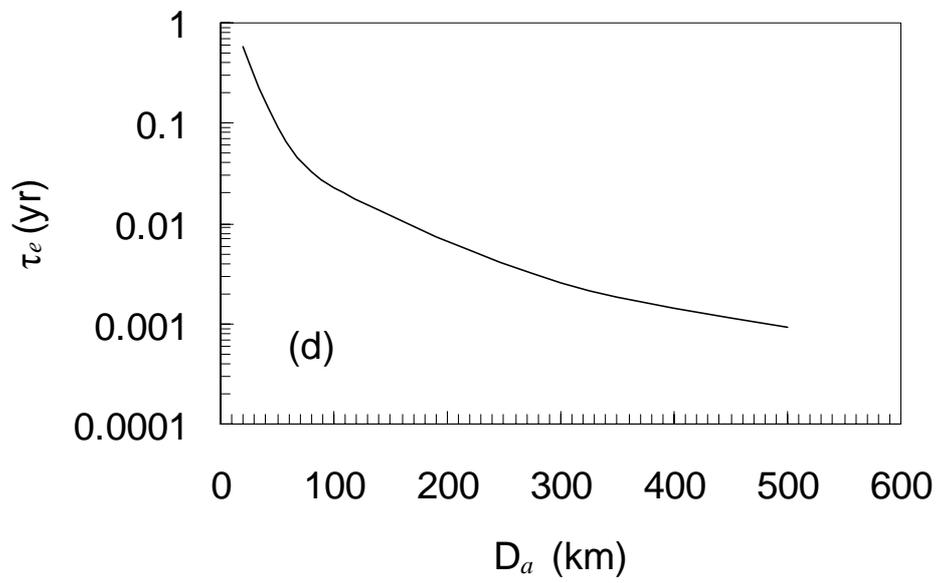

Fig 4. (c) The mean time interval between impacts versus projectiles of the given size or greater; (d) The expected time between impacts for asteroids of various sizes.

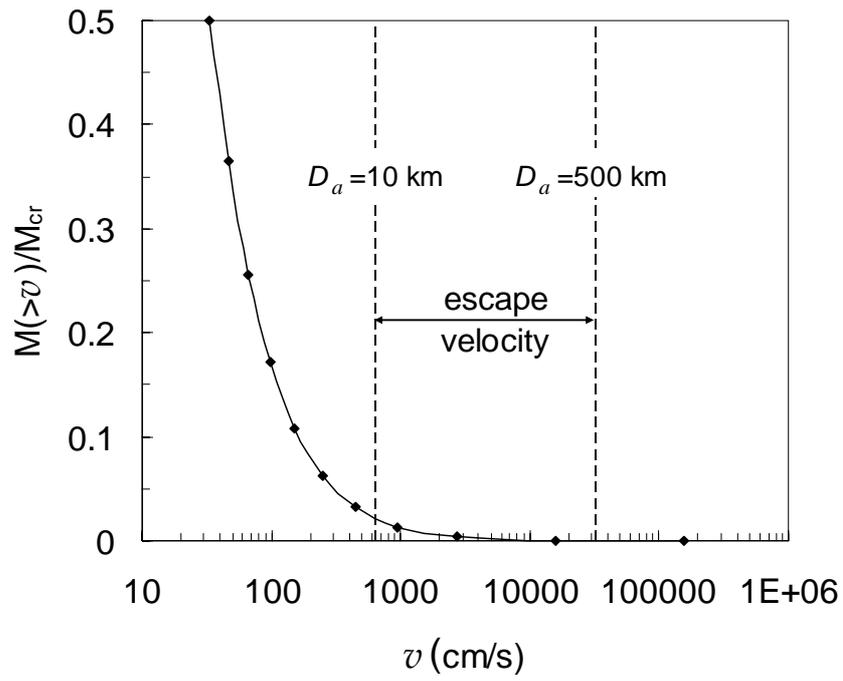

Fig. 5. Scaled mass of ejecta with velocity greater than $v$ against the ejection velocity. The range of an escape velocity for asteroids with diameter between 10 and 500 km is also shown.

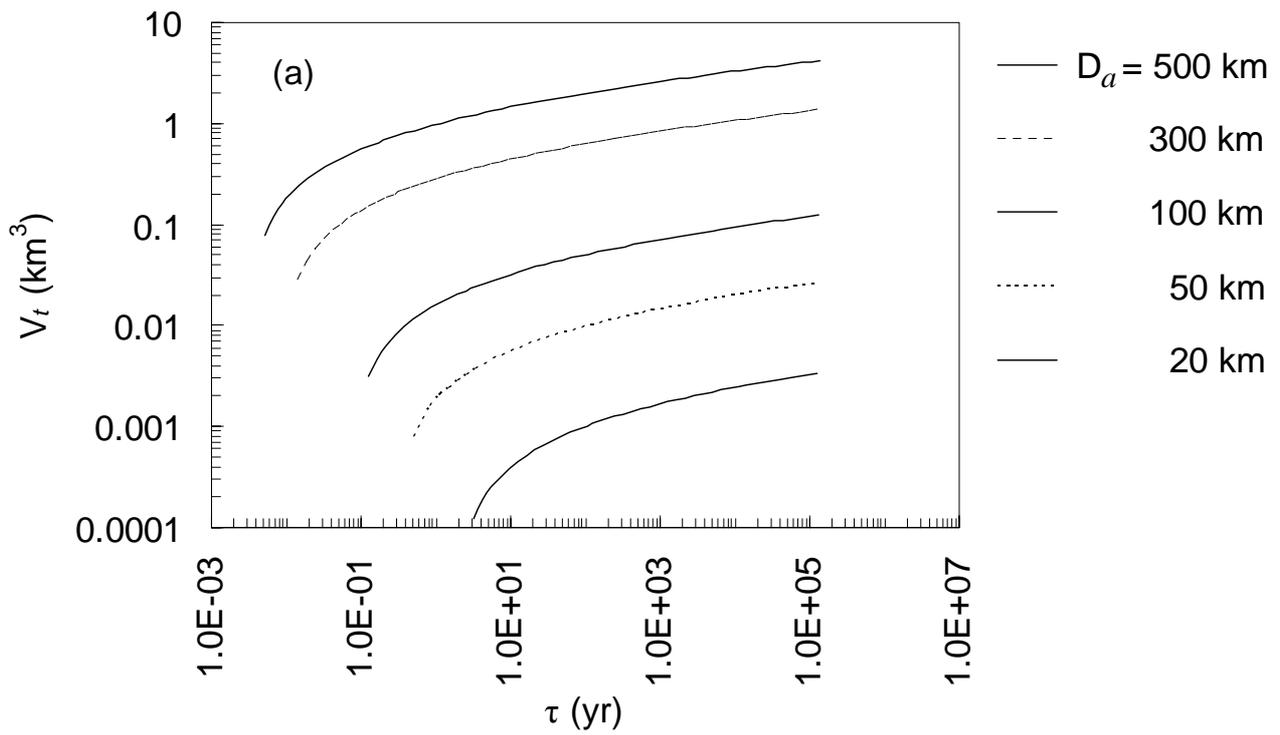

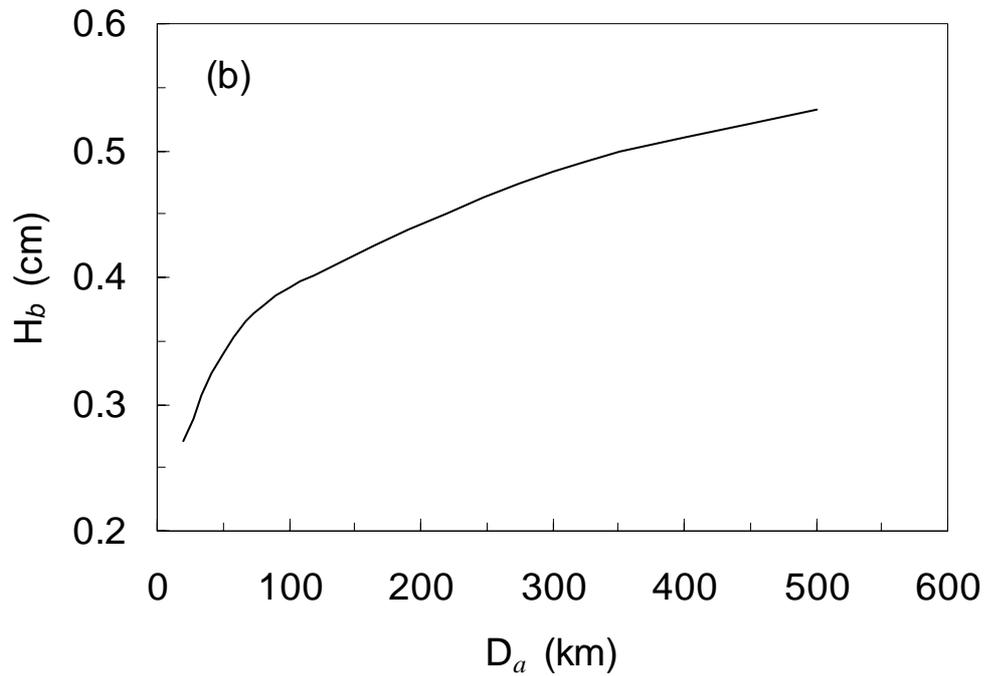

Fig. 6. (a) Volume of material ($V_j$) ejected during crater formation on asteroids of various diameter ($D_a$) as a function of the time period between impacts ($\tau$); (b) Average thickness ($H_b$) of the ejecta blanket formed on asteroids of various sizes in the time of $1.5 \times 10^5$ years;

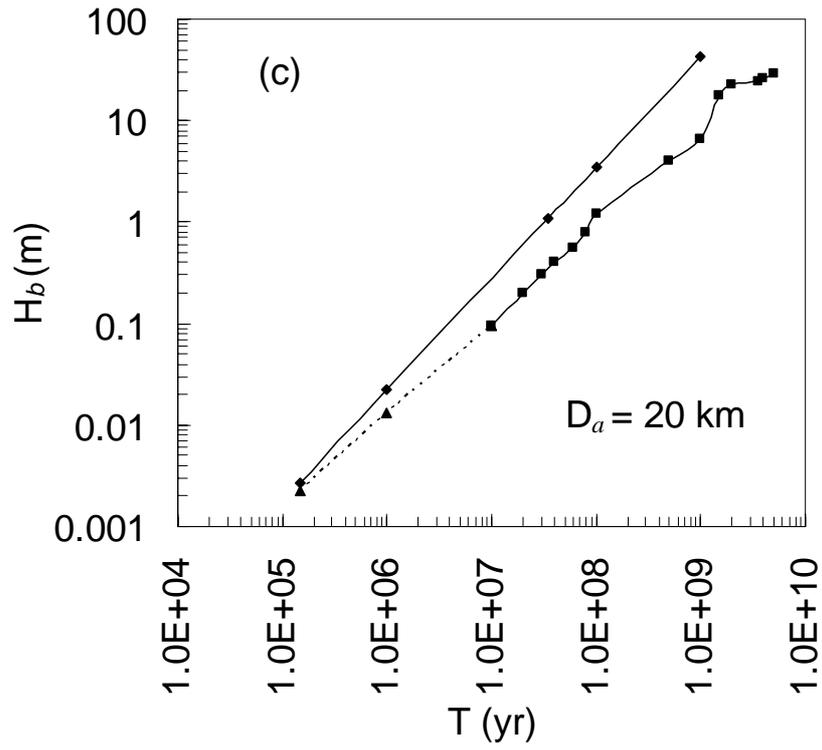

Fig. 6. (c) The growth of average regolith depth in an Eros-sized body over the bombardment time of ~ $10^9$ years. Diamonds, squares, and triangles denote, respectively, our calculations, the data from Richardson (2011), and extrapolation of these data towards shorter time span.

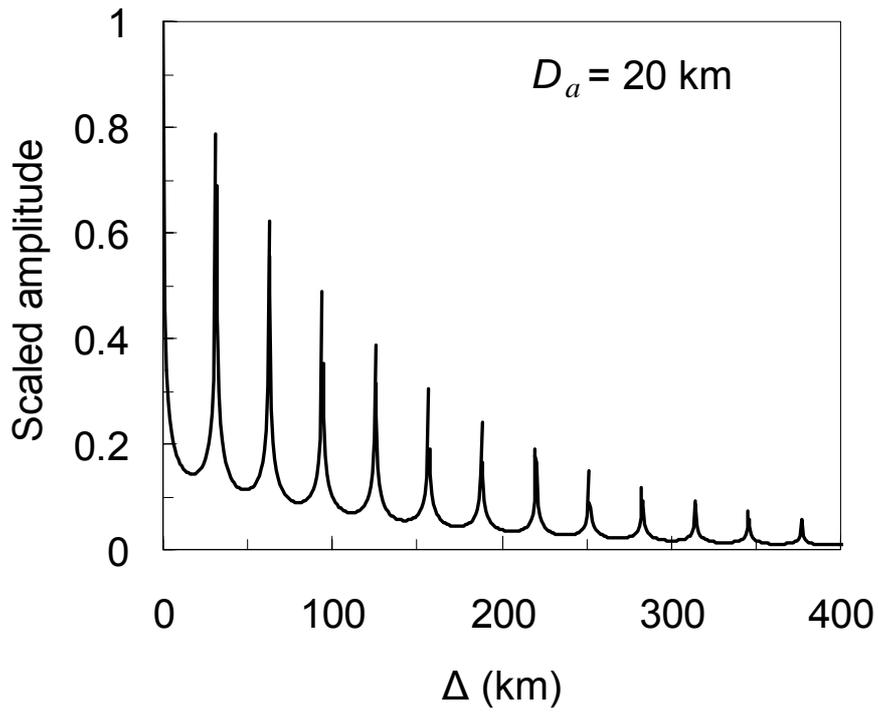

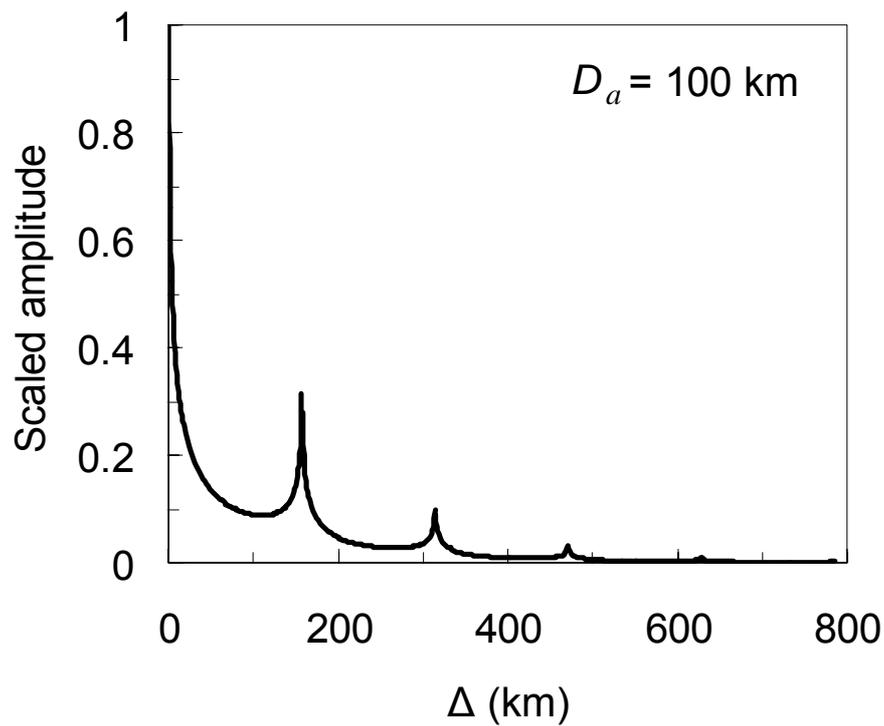

Fig. 7. The effect of multiple passages of a surface seismic wave between an impact site (the odd maxima) and its antipodal point (the even maxima) on spherical bodies with diameter of 20 and 100 km. Δ is distance between the impact site and front of the wave. Scaled amplitude of the surface wave equals 1 at Δ = 0.

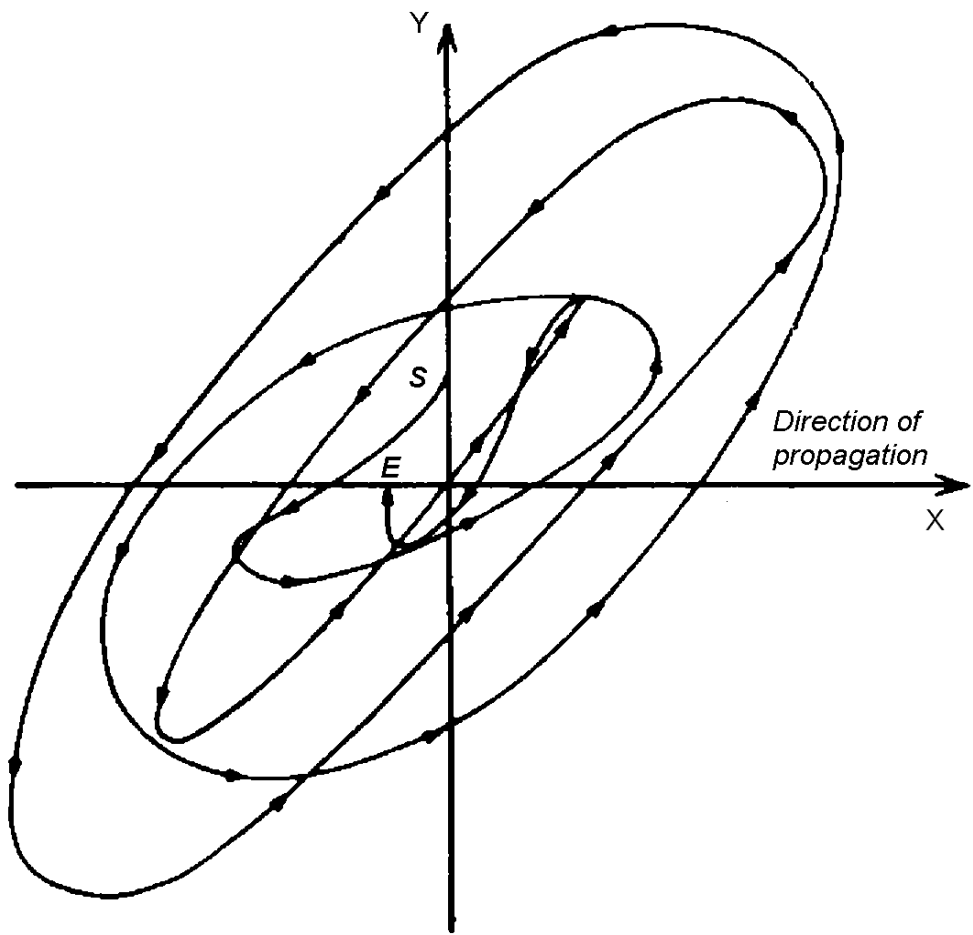

Fig. 8. Irregular motion of a particle on a terrestrial surface during passage of a Rayleigh waves. The waves propagate along the X-axis. *S* is the particle start position; *E* is the particle end position. This plot is fragment of Figure 2.13 from Sheriff and Geldart (1982).

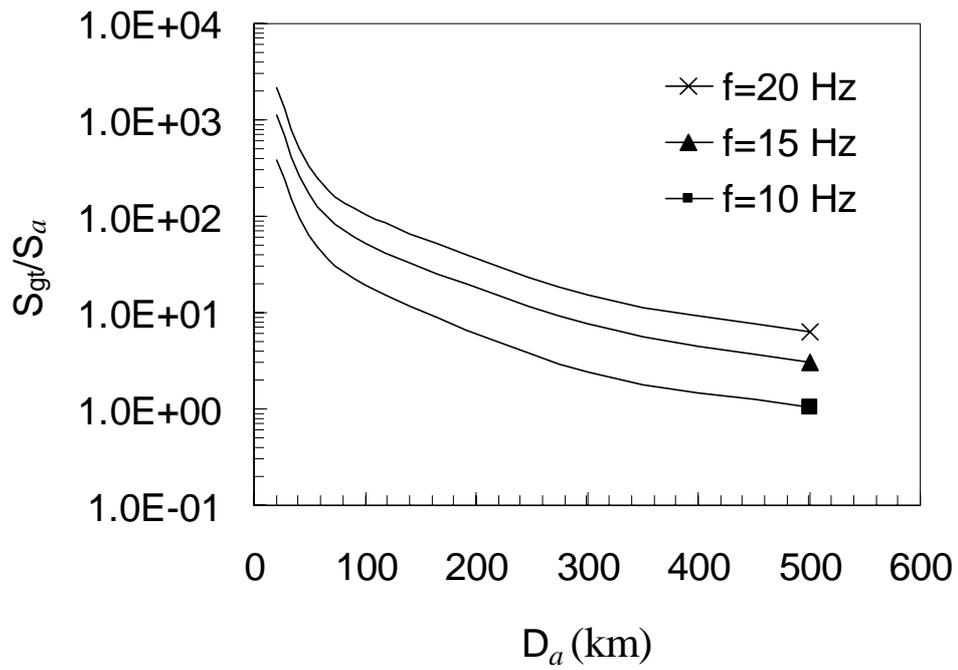

Fig. 9. Total area around an impact site ($S_{gt}$), where seismic-induced acceleration is greater than or equal to gravitational acceleration on an asteroid's surface as a function of the asteroid's diameter ($D_a$). These values were estimated for the time interval $T = 1.5 \; 10^5$ years and scaled to total area of the asteroid surfaces ($S_a$). $f$ is the primary seismic frequency of surface oscillations (Richardson et al., 2005).